\documentclass[letterpaper,10pt,conference]{ieeeconf}

\IEEEoverridecommandlockouts                              

\usepackage{microtype} 
\usepackage{amsmath,amssymb}
\usepackage{graphicx}
\usepackage{xcolor}
\usepackage{booktabs}
\usepackage{algorithm}
\usepackage{algpseudocode}
\usepackage{cite}
\usepackage{hyperref}

\algtext*{EndIf}
\algtext*{EndFor} 

\newtheorem{theorem}{Theorem}
\newtheorem{lemma}{Lemma}

\title{\LARGE \bf Control-Aware Separating Hyperplanes for Control Barrier Functions in Planar Navigation}

\author{Aadarsh~Ramachandiran and Rachel~Kalpana~Kalaimani%
\thanks{The authors are with the Department of Electrical Engineering, Indian Institute of Technology Madras, Chennai 600036, India (e-mail: ee23b001@smail.iitm.ac.in; rachel@ee.iitm.ac.in). (Corresponding author: Aadarsh Ramachandiran.)}%
\thanks{\protect\raggedright\urlstyle{same} Supplementary video (method intuition and baseline comparisons): \url{https://drive.google.com/file/d/1BrgA0pNK7Gj1Vc5Rr2ql40ZXmq7S6753/view}}%
}

\begin{document}

\maketitle
\thispagestyle{empty}
\pagestyle{empty}

\begin{abstract}
Separating hyperplanes are widely used in Control Barrier Function (CBF)-based collision avoidance, yet their orientations strongly govern control intervention---the deviation of the applied control from the robot's nominal command. Existing methods typically select hyperplane orientations from geometry alone. Jointly optimizing orientations and control yields a continuous, non-convex problem in which local nonlinear solvers suffer from initialization sensitivity, local minima, and unpredictable latency, while discretized orientation grids scale exponentially with obstacle count. For planar navigation of a polygonal robot among polygonal obstacles, we present a solver-free, control-aware separating hyperplane selection method that directly computes a global optimum of this joint problem at each step whenever it is feasible. Exploiting the geometry of Minkowski difference polygons, we reduce this continuous non-convex optimization to evaluating a polynomial-size set of candidate control interventions via closed-form algebraic checks, selecting the feasible candidate of minimum norm. Across 250 simulated navigation scenarios, our method reduces cumulative intervention cost by 3.6\% at the median and 27.3\% at the 90th percentile compared to a local nonlinear solver, and completion time by 0.6\% and 5.6\%, respectively. In scalability benchmarks, it achieves a 6.1$\times$ per-step speedup at 16 simultaneous obstacles.
\end{abstract}



\section{Introduction}
\label{sec:introduction}

Autonomous mobile robots (AMRs) deployed in warehouse intralogistics must navigate confined spaces while tracking their desired motion and avoiding collisions with dynamic obstacles~\cite{fragapane2021planning}. Control barrier functions (CBFs) enforce safety constraints while minimally modifying the nominal command~\cite{AmesEtAl2019CBFTheory, AmesEtAl2017CBFQP} with high-order CBFs (HOCBFs) extending this framework to higher relative degree systems, such as acceleration-controlled robots~\cite{NguyenSreenath2016ECBF, XiaoBelta2022HOCBF}.

Geometric safety constraints also depend on how robot and obstacle footprints are modeled. Simplified primitives such as points, circles, or ellipsoids~\cite{BorrmannEtAl2015SwarmCBF, FunadaEtAl2025RotatingHyperplanes, ThyriEtAl2020ASVCBF} inflate true footprints and add conservatism, whereas polygonal representations preserve shape fidelity for CBF-based collision avoidance~\cite{ThirugnanamZengSreenath2022PolytopeCBF, WuEtAl2025PolygonalCBF, WeiEtAl2026ConvexPrimitiveHOCBF}.

Separating hyperplanes provide a convenient means of enforcing collision avoidance between polygonal footprints~\cite{BorrmannEtAl2015SwarmCBF, FunadaEtAl2025RotatingHyperplanes, ThyriEtAl2020ASVCBF}. The choice of hyperplane orientation strongly governs how restrictive the resulting control constraint is. Existing approaches generally select this orientation using either closest-point methods, which align the normal vector along the line joining the nearest points of the robot and obstacle~\cite{LiuHuangBelta2026IterativePolytopeCBF}, or geometric optimization methods, which explicitly maximize spatial separation~\cite{BorrmannEtAl2015SwarmCBF, SanchezRonceroEtAl2025VO, GhaffariEtAl2018ZOHBarrier}. However, neither accounts for the robot's nominal command. Consequently, maximizing geometric separation does not necessarily keep small the resulting \emph{control intervention} $\Delta u = u - u_{\mathrm{nom}}$---the deviation of the applied control $u$ from the nominal command $u_{\mathrm{nom}}$ required to satisfy the safety constraint.

The least-restrictive hyperplane (LRH) CBF~\cite{trende2025lrh} jointly selects the separating hyperplane orientation and control input at each step to minimize the control intervention. However, this per-step joint problem is continuous and non-convex. Existing LRH implementations therefore rely on discretized orientation grids, which scale exponentially with obstacle count, or local nonlinear solvers, which suffer from initialization sensitivity, local minima, and unpredictable latency.

We present \textbf{Exact LRH (ELRH)}, which adopts the least-restrictive-hyperplane idea of~\cite{trende2025lrh} for polygonal footprints and solves the resulting joint hyperplane--control problem exactly, without numerical solvers, for planar navigation among dynamic polygonal obstacles. By exploiting the geometry of Minkowski difference polygons, ELRH reduces the continuous non-convex problem to evaluating a polynomial-size set of candidate control interventions via closed-form algebraic checks. This yields a bounded worst-case runtime without any numerical optimization solver, attractive for resource-constrained hardware, while provably recovering a global optimum of the per-step problem whenever it is feasible.

\textbf{Our main contributions are:}

\begin{itemize}
    \item \textbf{Joint Hyperplane-CBF and Control Formulation:} We address planar navigation among dynamic polygonal obstacles via a separating-hyperplane control barrier function, jointly selecting the hyperplane orientation and control input to minimize the norm of the control intervention $\Delta u$.

    \item \textbf{Solver-Free Method:} We develop ELRH, an algebraic procedure that constructs a polynomial-size set of candidate control interventions and selects the feasible candidate of minimum norm without numerical optimization.

    \item \textbf{Theoretical Optimality and Safety Guarantees:} We prove that evaluating this finite candidate set recovers a global optimum of the per-step problem whenever it is feasible, and derive an explicit time-step bound under which the chosen hyperplane and Zero-Order Hold (ZOH) control guarantee safety throughout each step.

    \item \textbf{Benchmark Validation:} Across 250 simulated dynamic navigation scenarios, ELRH achieves lower cumulative intervention cost and faster completion time than a local nonlinear solver, while runtime benchmarks confirm its scalability with increasing obstacle counts.
\end{itemize}

\section{Problem Formulation}
\label{sec:problem_formulation}

\subsection{System Dynamics and Geometry}

Consider a planar double-integrator robot with position and velocity $p, v \in \mathbb{R}^2$ and bounded control acceleration $u \in \mathcal{U} := \{u \in \mathbb{R}^2 : \|u\|_2 \leq u_{\max}\}$ (where $\|\cdot\|_2$ denotes the Euclidean norm):
\begin{equation}
\dot{p} = v, \qquad \dot{v} = u.
\end{equation}
The robot aims to track a nominal command $u_{\mathrm{nom}} \in \mathcal{U}$ while avoiding dynamic obstacles.

Both the robot and obstacles translate without rotation. The robot footprint $Q(p) \subset \mathbb{R}^2$ is a full-dimensional convex polygon defined by the convex hull $\operatorname{conv}\{\cdot\}$ of its $N_q$ body-frame vertex offsets $\{q_k^0\}_{k=1}^{N_q} \subset \mathbb{R}^2$: 
\begin{equation}
    Q(p) = \operatorname{conv}\{p + q_k^0\}_{k=1}^{N_q}.
\end{equation}

For each obstacle $i \in \{1, \ldots, N_{\mathrm{obs}}\}$, geometry is obtained from a map or online perception as $N_{z,i}$ boundary points $\{z_{i\ell}(t)\}_{\ell=1}^{N_{z,i}} \subset \mathbb{R}^2$. To enforce safety clearance $\delta_{\mathrm{clear}} > 0$, its raw convex hull is inflated via a Minkowski sum ($\oplus$) with the clearance square $\mathcal{C}_{\delta} = \{x \in \mathbb{R}^2 : \|x\|_\infty \leq \delta_{\mathrm{clear}}\}$ (where $\|\cdot\|_\infty$ denotes the infinity norm), yielding a convex polygon $\tilde{Z}_i(t)$ with $\tilde{N}_{z,i} \leq N_{z,i} + 4$ vertices:
\begin{equation}
    \tilde{Z}_i(t) = \operatorname{conv}\{z_{i\ell}(t)\}_{\ell=1}^{N_{z,i}} \oplus \mathcal{C}_{\delta} = \operatorname{conv}\{\tilde{z}_{im}(t)\}_{m=1}^{\tilde{N}_{z,i}}.
\end{equation}
Non-convex obstacles are partitioned into convex sub-polygons via standard polygon decomposition~\cite{HertelMehlhorn1985Decomposition, LienAmato2006ACD} and treated as independent convex obstacles; since the Minkowski sum distributes over set unions, inflating each sub-polygon is equivalent to inflating the original non-convex shape.

Under a discrete-time control loop with sampling period $\delta t$, the control input $u(t_k)$ is held constant over $t \in [t_k, t_k + \delta t)$ via ZOH, while each obstacle translates at a constant estimated velocity $v_i(t_k)$. Hereafter, we consider a single sampling interval and omit explicit time indexing where unambiguous.

\subsection{Hyperplane Separation and Barrier Construction}

For robot footprint $Q(p)$ and inflated obstacle polygon $\tilde{Z}_i(t)$, their signed separation along a non-zero normal vector $n \in \mathbb{R}^2 \setminus \{0\}$ pointing toward the robot is:
\begin{equation}
    h_i(p, t; n) = \min_{q \in Q(p)} n^\top q - \max_{z \in \tilde{Z}_i(t)} n^\top z = \min_{m \in \mathcal{I}_i} h_{i,m}(p, t; n),
    \label{eq:barrier}
\end{equation}
where $h_{i,m}(p, t; n) := n^\top r_{i,m}(p, t)$ is the vertex-wise barrier function for vertex $r_{i,m}(p, t)$ of the Minkowski difference polygon $M_i := Q(p) \oplus \left(-\tilde{Z}_i(t)\right)$, indexed by $\mathcal{I}_i := \{1, \ldots, |\mathcal{V}(M_i)|\}$, with $\mathcal{V}(\cdot)$ denoting the vertex set of a polygon and $\vert{}\cdot\vert{}$ its cardinality. A non-negative separation $h_i \ge 0$ (suppressing arguments where clear from context) guarantees that the interiors of $Q(p)$ and $\tilde{Z}_i(t)$ are disjoint (touching at most on their boundaries when $h_i = 0$), ensuring the robot remains separated from the uninflated obstacle by at least $\delta_{\mathrm{clear}}$ in the $\ell_\infty$ sense. Because $h_i = \min_{m \in \mathcal{I}_i} h_{i,m}$, requiring $h_i \ge 0$ is equivalent to enforcing $h_{i,m} \ge 0$ for every vertex:
\begin{equation}
n^\top r_{i,m} \ge 0, \quad \forall m \in \mathcal{I}_i. \label{eq:zero-cbf}
\end{equation}

With $n$ and $v_i$ held constant, the time derivatives of $h_{i,m}$ are $\dot{h}_{i,m} = n^\top v_{\mathrm{rel},i}$ and $\ddot{h}_{i,m} = n^\top u$, where $v_{\mathrm{rel},i} := v - v_i$. Safety is enforced via a relative-degree-2 HOCBF~\cite{XiaoBelta2022HOCBF} with linear class-$\mathcal{K}$ functions $\alpha_1(s) = \lambda_1 s$ and $\alpha_2(s) = \lambda_2 s$ ($\lambda_1, \lambda_2 > 0$). For each vertex $m \in \mathcal{I}_i$, the first-order condition $\psi_{i,m} := \dot{h}_{i,m} + \lambda_1 h_{i,m} \ge 0$ becomes:
\begin{equation}
n^\top\bigl(v_{\mathrm{rel},i} + \lambda_1 r_{i,m}\bigr) \ge 0, \quad \forall m \in \mathcal{I}_i, \label{eq:intermediate-cbf}
\end{equation}
and the second-order condition $\dot{\psi}_{i,m} + \lambda_2 \psi_{i,m} \ge 0$ yields:
\begin{equation}
n^\top u + (\lambda_1 + \lambda_2)n^\top v_{\mathrm{rel},i} + \lambda_1 \lambda_2 n^\top r_{i,m} \ge 0,\; \forall m \in \mathcal{I}_i. \label{eq:final-cbf}
\end{equation}
Admissible separating normals thus lie in $\mathcal{K}_i \setminus \{0\}$, where $\mathcal{K}_i := \{ n \in \mathbb{R}^2 : \text{\eqref{eq:zero-cbf} and \eqref{eq:intermediate-cbf} hold} \}$ is a cone. Since $h_i = \min_{m \in \mathcal{I}_i} h_{i,m}$ is a pointwise minimum, satisfying \eqref{eq:zero-cbf}--\eqref{eq:final-cbf} for all $m$ enforces the HOCBF conditions for $h_i$.

Finally, expressing \eqref{eq:final-cbf} in terms of control intervention $\Delta u := u - u_{\mathrm{nom}}$ yields the half-space constraint:
\begin{equation}
    n^\top \Delta u \ge \varphi_i(n),
    \label{eq:max-affine-cbf}
\end{equation}
where $\varphi_i(n) := \max_{m \in \mathcal{I}_i} n^\top w_{i,m}$ and $w_{i,m} := -u_{\mathrm{nom}} - (\lambda_1 + \lambda_2)v_{\mathrm{rel},i} - \lambda_1\lambda_2 r_{i,m}$.

\subsection{Joint Minimum-Intervention Problem}
\label{subsec:joint-problem}

We cast the least-restrictive-hyperplane idea of~\cite{trende2025lrh} as the following joint problem for polygonal footprints. At each sample step $t_k$, to track $u_{\mathrm{nom}}$ while guaranteeing safety across all dynamic obstacles ($i = 1, \ldots, N_{\mathrm{obs}}$), we jointly optimize the control intervention $\Delta u$ and separating normals $\{n_i\}_{i=1}^{N_{\mathrm{obs}}}$, minimizing the intervention cost $J(\Delta u) = \tfrac{1}{2}\|\Delta u\|_2^2$:
\begin{equation}
\begin{aligned}
\min_{\Delta u, \{n_i\}_{i=1}^{N_{\mathrm{obs}}}} \; & \frac{1}{2}\|\Delta u\|_2^2 \\
\text{s.t.} \; & u_{\mathrm{nom}} + \Delta u \in \mathcal{U}, \\
& n_i \in \mathcal{K}_i \setminus \{0\}, \\
& n_i^\top \Delta u \ge \varphi_i(n_i).
\end{aligned}
\label{eq:joint-problem}
\end{equation}
In this sense, a hyperplane is least restrictive when it admits the smallest intervention cost. The resulting control $u(t_k) = u_{\mathrm{nom}}(t_k) + \Delta u^*$ is held constant over the ZOH interval $[t_k, t_k + \delta t)$ (ensuring safety throughout the interval; analyzed in Section~\ref{subsec:complexity}). Because the constraints are positively homogeneous in $n_i$, any non-zero normal can be normalized to unit length. However, the bilinear coupling between $\Delta u$ and $n_i$ in \eqref{eq:max-affine-cbf} renders \eqref{eq:joint-problem} non-convex, typically requiring nonlinear solvers or discrete orientation grid searches.


\section{Methodology}
\label{sec:method}
Although \eqref{eq:joint-problem} is non-convex, we solve it exactly by reducing each obstacle constraint to a finite union of half-spaces, constructing a polynomial-size candidate set that contains a global optimum (if it is feasible), and analyzing complexity and ZOH safety.


\subsection{Finite Half-Space Representation of Obstacle Constraints}
\label{subsec:halfspace-representation}

A control intervention $\Delta u$ is safe with respect to obstacle $i$ if and only if it belongs to the feasible set:
\begin{equation}
    \mathcal{F}_i := \bigl\{ \Delta u \in \mathbb{R}^2 : \exists n \in \mathcal{K}_i \setminus \{0\} \text{ s.t. \eqref{eq:max-affine-cbf}} \bigr\}.
    \label{eq:single-obstacle-set}
\end{equation}

To decompose the piecewise-linear function $\varphi_i(n)$ into linear pieces, we partition $\mathcal{K}_i$ into sub-cones $\mathcal{C}_{i,m}$ where the $m$-th term attains the maximum:
\begin{equation}
\mathcal{C}_{i,m} := \bigl\{ n \in \mathcal{K}_i : n^\top(w_{i,m} - w_{i,k}) \ge 0,\; \forall k \in \mathcal{I}_i \bigr\}.
\label{eq:active-regions}
\end{equation}
By construction, $\mathcal{K}_i$ is a polygonal cone defined by finitely many homogeneous linear inequalities, and because the full-dimensional polygon $M_i$ has vertices spanning $\mathbb{R}^2$, we have $\mathcal{K}_i \cap (-\mathcal{K}_i) = \{0\}$; hence, $\mathcal{K}_i$ is pointed. Since each sub-cone $\mathcal{C}_{i,m} \subseteq \mathcal{K}_i$ is formed by intersecting $\mathcal{K}_i$ with additional homogeneous linear inequalities, it is likewise a pointed polygonal cone and is therefore generated by its finite set of unit-length extreme-ray generators, denoted $\widehat{\operatorname{Ext}}(\mathcal{C}_{i,m})$.

On each $\mathcal{C}_{i,m}$, $\varphi_i(n) = n^\top w_{i,m}$ is linear. Any non-zero normal $n \in \mathcal{C}_{i,m} \setminus \{0\}$ can therefore be written as a conic combination $n = \sum_j \alpha_j r_j$ with $r_j \in \widehat{\operatorname{Ext}}(\mathcal{C}_{i,m})$ and $\alpha_j \ge 0$ (not all zero). Substituting this expansion into \eqref{eq:max-affine-cbf} yields $\sum_j \alpha_j \bigl(r_j^\top \Delta u - \varphi_i(r_j)\bigr) \ge 0$. Since $\alpha_j \ge 0$ with at least one coefficient strictly positive, there exists $n \in \mathcal{C}_{i,m} \setminus \{0\}$ satisfying \eqref{eq:max-affine-cbf} if and only if $r_j^\top \Delta u \ge \varphi_i(r_j)$ holds for at least one generator $r_j \in \widehat{\operatorname{Ext}}(\mathcal{C}_{i,m})$.

Collecting the unit-length extreme-ray generators across all sub-cones yields the finite candidate direction set:
\begin{equation}
\mathcal{R}_i := \bigcup_{m \in \mathcal{I}_i} \widehat{\operatorname{Ext}}(\mathcal{C}_{i,m}).
\label{eq:candidate-ray-set}
\end{equation}
For each $r \in \mathcal{R}_i$, let $H_{i,r} := \{\Delta u \in \mathbb{R}^2 : r^\top \Delta u \ge \varphi_i(r)\}$ denote the linear half-space defined by \eqref{eq:max-affine-cbf} at $n = r$. The single-obstacle feasible set $\mathcal{F}_i$ in \eqref{eq:single-obstacle-set} then reduces to the finite union:
\begin{equation}
\mathcal{F}_i = \bigcup_{r \in \mathcal{R}_i} H_{i,r}.
\label{eq:admissible-halfspace-union}
\end{equation}

\subsection{Explicit Construction of Candidate Normals}
\label{subsec:ray-enumeration}

To enumerate the extreme rays of $\mathcal{C}_{i,m}$, we examine the rays along which its defining inequalities hold with equality---arising either from the definition of $\mathcal{K}_i$ or from \eqref{eq:active-regions}. The former yield the boundary rays of $\mathcal{K}_i$. The latter occur when two distinct vertices $r_{i,a}, r_{i,b}$ of $M_i$ attain the same maximum in $\varphi_i(n)$, i.e., $n^\top(w_{i,a}-w_{i,b})=0$. Since $w_{i,a}-w_{i,b}=-\lambda_1\lambda_2(r_{i,a}-r_{i,b})$ with $\lambda_1,\lambda_2 > 0$, any tying normal $n$ must be orthogonal to $e = r_{i,b} - r_{i,a}$. Because $M_i$ is convex, the extreme rays generated by such ties correspond to adjacent vertices forming boundary edges of $M_i$; ordering the vertices of $M_i$ counterclockwise, each boundary edge $e = r_{i,b} - r_{i,a}$ has inward unit normal $n = \mathbf{J}e/\|e\|_2$, where $\mathbf{J} = \bigl[\begin{smallmatrix}0 & -1 \\ 1 & 0\end{smallmatrix}\bigr]$ is the $90^\circ$ counterclockwise rotation matrix. Combining the unit boundary rays of $\mathcal{K}_i$ with these inward unit edge normals restricted to $\mathcal{K}_i$ yields $\mathcal{R}_i$, whose cardinality is bounded as follows.

\begin{lemma}[Candidate Set Cardinality]
\label{lem:direction-count}
For convex polygonal footprints $Q(p)$ and $\tilde{Z}_i(t)$ with $N_q$ and $\tilde{N}_{z,i}$ vertices, respectively, the candidate direction set $\mathcal{R}_i$ in \eqref{eq:candidate-ray-set} satisfies
\begin{equation}
    |\mathcal{R}_i| \le N_q + \tilde{N}_{z,i} + 2.
    \label{eq:direction-count}
\end{equation}
\end{lemma}

\begin{proof}
$\mathcal{K}_i$ contributes at most $2$ boundary rays. As the Minkowski sum of polygons with $N_q$ and $\tilde{N}_{z,i}$ vertices, $M_i$ has at most $N_q+\tilde{N}_{z,i}$ edges, each giving at most one inward unit normal.
\end{proof}


\subsection{Joint Feasibility and Global Optimality}
\label{subsec:joint-feasibility}

Intersecting $\Delta\mathcal{U}:=\{\Delta u\in\mathbb{R}^2 : \|u_{\mathrm{nom}}+\Delta u\|_2\le u_{\max}\}$ with \eqref{eq:admissible-halfspace-union} across all $N_{\mathrm{obs}}$ obstacles yields the joint feasible set:
\begin{equation}
\mathcal{F}_{\mathrm{joint}}=\Delta\mathcal{U}\cap\bigcap_{i=1}^{N_{\mathrm{obs}}}\bigcup_{r\in\mathcal{R}_i}H_{i,r}.
\label{eq:joint-feasible-set}
\end{equation}
Selecting one direction $r_i\in\mathcal{R}_i$ per obstacle defines a closed convex region $\mathcal{P}(r_1,\ldots,r_{N_{\mathrm{obs}}}):=\Delta\mathcal{U}\cap\bigcap_{i=1}^{N_{\mathrm{obs}}} H_{i,r_i}$. Expanding \eqref{eq:joint-feasible-set} over all tuples $(r_1,\ldots,r_{N_{\mathrm{obs}}})$ expresses $\mathcal{F}_{\mathrm{joint}}$ as a union of $\prod_{i=1}^{N_{\mathrm{obs}}}|\mathcal{R}_i|$ convex regions, which is exponential in $N_{\mathrm{obs}}$ and intractable to enumerate directly.

In the planar control space, however, a minimum-norm intervention over any $\mathcal{P}(r_1,\ldots,r_{N_{\mathrm{obs}}})$ is determined by at most two active boundary constraints, reducing the global search to a finite set of analytic candidates. For each $r\in\mathcal{R}_i$ (with $\|r\|_2=1$), let $\beta_{i,r}:=\varphi_i(r)$ and $\ell_{i,r}:=\{\Delta u\in\mathbb{R}^2 : r^\top\Delta u=\beta_{i,r}\}$. We construct the candidate set $\mathcal{C}=\bigcup_{k=0}^3\mathcal{C}_k$ as follows:
\begin{itemize}
    \item \textbf{Origin ($\mathcal{C}_0$):} The zero intervention $\mathcal{C}_0 := \{0\}$.

    \item \textbf{Single-obstacle projection ($\mathcal{C}_1$):} For $\beta_{i,r} > 0$, the orthogonal projection of the origin onto $\ell_{i,r}$ is:
    \begin{equation}
        \Delta u_{i,r}^{\mathrm{foot}} = \beta_{i,r} r.
        \label{eq:foot-candidate}
    \end{equation}

    \item \textbf{Two-obstacle intersection ($\mathcal{C}_2$):} For $i < j$ and linearly independent $r \in \mathcal{R}_i, s \in \mathcal{R}_j$, the line intersection is:
    \begin{equation}
        \Delta u_{i,j,r,s}^{\mathrm{int}} = \begin{bmatrix} r^\top \\ s^\top \end{bmatrix}^{-1} \begin{bmatrix} \beta_{i,r} \\ \beta_{j,s} \end{bmatrix}.
        \label{eq:line-intersection}
    \end{equation}
    We retain intersections admitting a representation $\Delta u_{i,j,r,s}^{\mathrm{int}} = c_1 r + c_2 s$ with $c_1, c_2 > 0$.

    \item \textbf{Obstacle--actuator intersection ($\mathcal{C}_3$):} Decomposing $\Delta u = \beta_{i,r} r + \eta \mathbf{J}r$ in the orthonormal basis $\{r, \mathbf{J}r\}$, the intersections of $\ell_{i,r}$ with the actuator boundary $\|u_{\mathrm{nom}} + \Delta u\|_2 = u_{\max}$ are:
    \begin{equation}
    \begin{aligned}
    \Delta u_{i,r}^{\mathrm{circ},\pm} &= \beta_{i,r} r + \eta_{i,r}^{\pm} \mathbf{J}r, \\
    \eta_{i,r}^{\pm} &= -(\mathbf{J}r)^\top u_{\mathrm{nom}} \\
    &\quad \pm \sqrt{u_{\max}^2 - (\beta_{i,r} + r^\top u_{\mathrm{nom}})^2},
    \end{aligned}
    \label{eq:circle-intersections}
    \end{equation}
    evaluated whenever the radicand is non-negative.
\end{itemize}

A candidate $\Delta u \in \mathcal{C}$ is admissible if it satisfies all actuator and obstacle constraints:
\begin{equation}
\begin{split}
& \|u_{\mathrm{nom}} + \Delta u\|_2 \le u_{\max}, \quad \text{and} \\
& \forall i \in \{1, \ldots, N_{\mathrm{obs}}\},\ \exists r \in \mathcal{R}_i :\ r^\top \Delta u \ge \beta_{i,r}.
\end{split}
\label{eq:acceptance-test}
\end{equation}

\begin{theorem}[Global Optimality]
\label{thm:global-optimality}
If $\mathcal{F}_{\mathrm{joint}} \neq \varnothing$ and $0 \in \Delta\mathcal{U}$, every global minimizer $\Delta u^\star \in$ $\arg\min_{\Delta u \in \mathcal{F}_{\mathrm{joint}}} \frac{1}{2}\|\Delta u\|_2^2$ belongs to $\mathcal{C}=\bigcup_{k=0}^3\mathcal{C}_k$. Consequently, selecting a feasible candidate in $\mathcal{C}$ with minimum norm yields a globally optimal control intervention.
\end{theorem}

\begin{proof}
Since $\mathcal{F}_{\mathrm{joint}}$ is a non-empty, closed subset of the compact disk $\Delta\mathcal{U}$, a global minimizer $\Delta u^\star$ exists by the Extreme Value Theorem. Let $\mathcal{P} = \Delta\mathcal{U} \cap \bigcap_{i=1}^{N_{\mathrm{obs}}} H_{i,r_i}$ be a convex region in the finite union \eqref{eq:joint-feasible-set} containing $\Delta u^\star$. Because $\mathcal{P} \subseteq \mathcal{F}_{\mathrm{joint}}$, $\Delta u^\star$ also minimizes $\|\Delta u\|_2$ over $\mathcal{P}$.

If $\Delta u^\star = 0$, then $\Delta u^\star \in \mathcal{C}_0$. Assume henceforth that $\Delta u^\star \neq 0$.

If the actuator constraint is active ($\|u_{\mathrm{nom}} + \Delta u^\star\|_2 = u_{\max}$), at least one obstacle constraint in $\mathcal{P}$ must also be active; otherwise, because $\Delta\mathcal{U}$ is convex and $0 \in \Delta\mathcal{U}$, moving $\Delta u^\star$ slightly along the line segment toward $0$ would strictly decrease $\|\Delta u\|_2$ while remaining in $\Delta\mathcal{U}$ and strictly satisfying all inactive obstacle constraints. Thus, $\Delta u^\star$ lies at the intersection of an active obstacle boundary line $\ell_{i,r_i}$ and the actuator boundary, placing it in $\mathcal{C}_3$.

Now suppose the actuator constraint is inactive ($\|u_{\mathrm{nom}} + \Delta u^\star\|_2 < u_{\max}$), so $\Delta u^\star \in \operatorname{int}(\Delta\mathcal{U})$ (with $\operatorname{int}(\cdot)$ denoting the set interior) and $\mathcal{P}$ is locally polyhedral around $\Delta u^\star$. Let $\mathcal{A} \subseteq \{1,\ldots,N_{\mathrm{obs}}\}$ denote the index set of active obstacle constraints in $\mathcal{P}$. The tangent cone to $\mathcal{P}$ at $\Delta u^\star$ is $T_{\mathcal{P}}(\Delta u^\star) = \{d \in \mathbb{R}^2 : r_i^\top d \ge 0,\; \forall i \in \mathcal{A}\}$. First-order optimality requires $(\Delta u^\star)^\top d \ge 0$ for all $d \in T_{\mathcal{P}}(\Delta u^\star)$, which by Farkas' lemma (dual-cone characterization) implies $\Delta u^\star \in \operatorname{cone}(\{r_i : i \in \mathcal{A}\})$.

By conic Carath\'eodory's theorem in $\mathbb{R}^2$, $\Delta u^\star$ is a positive linear combination of at most two linearly independent active normals (any dependent pair of unit normals reduces to a single positive multiple since $\Delta u^\star \neq 0$). If one normal $r_i$ suffices, $\Delta u^\star = c r_i$ for some $c > 0$; substituting into $r_i^\top \Delta u^\star = \beta_{i,r_i}$ with $\|r_i\|_2 = 1$ yields $c = \beta_{i,r_i} > 0$, so $\Delta u^\star = \beta_{i,r_i} r_i \in \mathcal{C}_1$. Otherwise, two linearly independent active normals $r_i, r_j$ ($i < j$) yield $\Delta u^\star = c_1 r_i + c_2 r_j$ with $c_1, c_2 > 0$; since both constraints are active, $\Delta u^\star$ is the unique intersection of $\ell_{i,r_i}$ and $\ell_{j,r_j}$, placing it in $\mathcal{C}_2$.

Hence, every global minimizer belongs to $\mathcal{C}$, and minimizing the objective over all feasible candidates in $\mathcal{C}$ recovers a global minimizer.
\end{proof}

\begin{algorithm}[!t]
\caption{ELRH Online Control Loop}
\label{alg:lrh_cbf}
\small
\begin{algorithmic}[1]
\Require Robot state $(p, v)$, inflated obstacles $\{\tilde{Z}_i(t_k), v_i\}_{i=1}^{N_{\mathrm{obs}}}$, nominal command $u_{\mathrm{nom}}$
\Ensure Safe control input $u^\star$ on $[t_k, t_k+\delta t)$, or \textbf{Infeasible}
\For{each obstacle $i = 1, \ldots, N_{\mathrm{obs}}$}
    \State $M_i \gets Q(p) \oplus (-\tilde{Z}_i(t_k))$
    \State Construct $\mathcal{R}_i$ via~\eqref{eq:candidate-ray-set} and compute $\beta_{i,r} \gets \varphi_i(r) \ \forall r \in \mathcal{R}_i$
    \If{$\mathcal{R}_i = \varnothing$}
        \State \Return \textbf{Infeasible}
    \EndIf
\EndFor
\State Assemble candidate set $\mathcal{C} \gets \bigcup_{k=0}^3 \mathcal{C}_k$
\For{each $\Delta u \in \mathcal{C}$ in ascending order of $\|\Delta u\|_2^2$}
    \If{$\Delta u$ satisfies~\eqref{eq:acceptance-test}}
        \State \Return $u^\star \gets u_{\mathrm{nom}} + \Delta u$
    \EndIf
\EndFor
\State \Return \textbf{Infeasible}
\end{algorithmic}
\end{algorithm}

\subsection{Computational Complexity and Sampled-Data Invariance}
\label{subsec:complexity}

Let $V_i := |\mathcal{R}_i|$ and $V := \max_i V_i$. Precomputing the candidate directions $\mathcal{R}_i$ and offsets $\beta_{i,r}$ across all obstacles requires $O(N_{\mathrm{obs}} V^2)$ operations. Counting the origin ($1$), single-obstacle projections ($\le N_{\mathrm{obs}} V$), pairwise obstacle intersections ($\le \binom{N_{\mathrm{obs}}}{2} V^2$), and obstacle--actuator intersections ($\le 2N_{\mathrm{obs}} V$) bounds the candidate set size by
\begin{equation}
    |\mathcal{C}| \le 1 + 3N_{\mathrm{obs}}V + \binom{N_{\mathrm{obs}}}{2}V^2 = O(N_{\mathrm{obs}}^2 V^2).
    \label{eq:candidate-count}
\end{equation}
Verifying whether a candidate belongs to $\mathcal{F}_{\mathrm{joint}}$ in \eqref{eq:joint-feasible-set} requires checking $\Delta u \in \Delta\mathcal{U}$ in $O(1)$ time and testing $r^\top \Delta u \ge \beta_{i,r}$ for at most $V$ directions per obstacle, costing $O(N_{\mathrm{obs}} V)$ per candidate. The overall procedure therefore yields a worst-case runtime of $O(N_{\mathrm{obs}}^3 V^3)$. By Lemma~\ref{lem:direction-count}, $V \le N_q + \max_i \tilde{N}_{z,i} + 2$, so selecting a globally optimal control intervention runs in polynomial time with respect to the obstacle count and polygon vertex counts. We next show that one-step forward invariance holds across the entire inter-sample interval $[t_k, t_k + \delta t)$ under ZOH, provided $\delta t$ satisfies an explicit upper bound.

\begin{theorem}[One-Step Sampled-Data Invariance]
\label{thm:forward-invariance}
For each obstacle $i \in \{1, \ldots, N_{\mathrm{obs}}\}$, let $n_{i,k} := n_i(t_k)$, $h_{i,m}(t) := n_{i,k}^\top r_{i,m}(p(t), t)$ for $m \in \mathcal{I}_i$, and $h_i(t) := \min_{m \in \mathcal{I}_i} h_{i,m}(t)$. Define
\begin{equation}
\begin{split}
\psi_{i,m}(t) &:= \dot h_{i,m}(t) + \lambda_1 h_{i,m}(t), \\
\rho_{i,m}(t) &:= \dot \psi_{i,m}(t) + \lambda_2 \psi_{i,m}(t),
\end{split}
\end{equation}
where $\lambda_1, \lambda_2 > 0$. Suppose over $[t_k, t_k + \delta t)$: (i)~$n_{i,k}$ and $u^\star = u_{\mathrm{nom}} + \Delta u^\star$ are held constant via ZOH; (ii)~$\dot v_i = 0$; and (iii)~at $t_k$,
\begin{equation}
    h_{i,m}(t_k) \ge 0, \quad \psi_{i,m}(t_k) \ge 0, \quad \rho_{i,m}(t_k) \ge 0, \quad \forall m \in \mathcal{I}_i.
    \label{eq:initial-hocbf-conds}
\end{equation}
Then $h_i(t) \ge 0$ for all $t \in [t_k, t_k + \delta t)$, and $\dot h_i(t) + \lambda_1 h_i(t) \ge 0$ almost everywhere on $[t_k, t_k + \delta t)$, provided the sampling step $\delta t$ satisfies
\begin{equation}
\delta t \le \frac{\sqrt{\lambda_2^2 + 2\lambda_1(\lambda_1 + \lambda_2)} - \lambda_2}{\lambda_1(\lambda_1 + \lambda_2)}.
\label{eq:dt-hocbf-bound}
\end{equation}
\end{theorem}

\begin{proof}
Let $h_{i,m,k} := h_{i,m}(t_k)$, $\dot h_{i,m,k} := \dot h_{i,m}(t_k)$, $\ddot h_{i,m,k} := \ddot h_{i,m}(t_k)$, $\psi_{i,m,k} := \psi_{i,m}(t_k)$, and $\rho_{i,m,k} := \rho_{i,m}(t_k)$. Condition~(iii) holds by construction: for the selected $n_{i,k}\in\mathcal{K}_i$, \eqref{eq:zero-cbf}, \eqref{eq:intermediate-cbf} and \eqref{eq:max-affine-cbf} (equivalently \eqref{eq:final-cbf}) state $h_{i,m,k}\ge0$, $\psi_{i,m,k}\ge0$ and $\rho_{i,m,k}\ge0$. Under ZOH with $\dot v_i = 0$, the relative acceleration $\ddot r_{i,m} = u^\star$ is constant over $s \in [0, \delta t)$, yielding a constant scalar acceleration $\ddot h_{i,m} = n_{i,k}^\top u^\star$. Consequently, the exact second-order Taylor expansion of $\psi_{i,m}(t_k+s)$ on $[0, \delta t)$ is
\begin{equation}
    \psi_{i,m}(t_k+s) = \psi_{i,m,k} + (\ddot h_{i,m,k} + \lambda_1\dot h_{i,m,k})s + \frac{1}{2}\lambda_1\ddot h_{i,m,k}s^2.
\end{equation}
Using $\ddot h_{i,m,k} + \lambda_1\dot h_{i,m,k} = \rho_{i,m,k} - \lambda_2\psi_{i,m,k}$ and $\ddot h_{i,m,k} = \rho_{i,m,k} - (\lambda_1 + \lambda_2)\psi_{i,m,k} + \lambda_1^2 h_{i,m,k}$ rearranges this expression to
\begin{equation}
\begin{split}
\psi_{i,m}(t_k+s) ={}& \psi_{i,m,k}\Bigl[1 - \lambda_2 s - \tfrac{1}{2}\lambda_1(\lambda_1 + \lambda_2)s^2\Bigr] \\
& + \tfrac{1}{2}\lambda_1^3 h_{i,m,k} s^2 + \rho_{i,m,k}\Bigl(s + \tfrac{1}{2}\lambda_1 s^2\Bigr).
\end{split}
\label{eq:psi-expansion}
\end{equation}
By assumption~(iii), $h_{i,m,k} \ge 0$, $\psi_{i,m,k} \ge 0$, and $\rho_{i,m,k} \ge 0$. Bound~\eqref{eq:dt-hocbf-bound} is the positive root of $1 - \lambda_2 s - \frac{1}{2}\lambda_1(\lambda_1 + \lambda_2)s^2 = 0$, ensuring that the bracketed quadratic is non-negative for all $s \in [0, \delta t)$. Since the remaining coefficients $\frac{1}{2}\lambda_1^3 s^2$ and $s + \frac{1}{2}\lambda_1 s^2$ are non-negative for $s \ge 0$, it follows that $\psi_{i,m}(t_k+s) \ge 0$ for all $s \in [0, \delta t)$.

Integrating $\dot h_{i,m}(t) + \lambda_1 h_{i,m}(t) = \psi_{i,m}(t)$ yields
\begin{equation}
    h_{i,m}(t_k+s) = e^{-\lambda_1 s} h_{i,m,k} + \int_0^s e^{-\lambda_1(s-\xi)} \psi_{i,m}(t_k+\xi)\,d\xi,
\end{equation}
which is non-negative for all $s \in [0, \delta t)$ since $h_{i,m,k} \ge 0$ and $\psi_{i,m} \ge 0$ for every $m \in \mathcal{I}_i$.

Finally, $h_i(t) = \min_{m \in \mathcal{I}_i} h_{i,m}(t) \ge 0$ holds pointwise on $[t_k, t_k + \delta t)$. At every $t \in (t_k, t_k + \delta t)$ where $h_i$ is differentiable, there exists an active index $m^\star(t) \in \arg\min_{m \in \mathcal{I}_i} h_{i,m}(t)$ such that $\dot h_i(t) = \dot h_{i,m^\star(t)}(t)$, which gives $\dot h_i(t) + \lambda_1 h_i(t) = \psi_{i,m^\star(t)}(t) \ge 0$ almost everywhere on $[t_k, t_k + \delta t)$.
\end{proof}

\subsection{Online Control Algorithm}
\label{subsec:execution}

The discrete-time control loop executes at each sampling instant $t_k$, as summarized in Algorithm~\ref{alg:lrh_cbf}. Sorting the generated candidates in $\mathcal{C}$ in ascending order of $\|\Delta u\|_2^2$ and terminating at the first feasible point recovers a global minimizer while avoiding redundant feasibility checks in practice. If no admissible normal exists or no candidate is feasible, the algorithm returns an infeasible status, and a backup controller is required.

\begin{figure}[!t]
    \centering
    \includegraphics[width=\columnwidth]{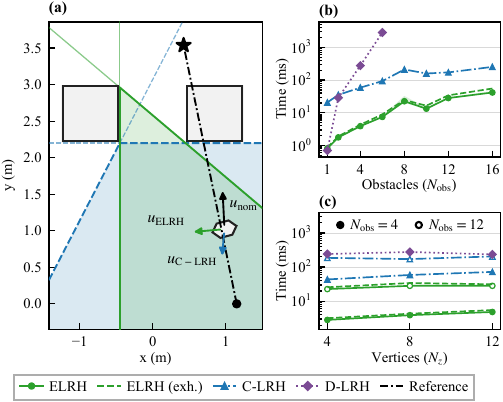}
\caption{Performance and scalability of ELRH. (a) Static doorway: ELRH selects the globally optimal half-space (green); C-LRH (continuous, IPOPT-based LRH) converges to a restrictive local minimum (blue). (b) Median solve time vs.\ $N_{\mathrm{obs}}$ ($N_z=8$) for ELRH, ELRH with exhaustive search (exh.), C-LRH and D-LRH. (c) Median solve time vs.\ footprint vertex count $N_z$.}
\label{fig:figure}
\end{figure}

\section{Experiments and Results}
\label{sec:experiments}

We evaluate ELRH on three questions: \textit{(i)}~does control-aware hyperplane selection improve closed-loop navigation; \textit{(ii)}~how much does exact global optimality outperform nonlinear solvers prone to local minima; and \textit{(iii)}~how does the candidate search scale with obstacle count and polygon complexity? All simulations use semi-implicit Euler integration ($\delta t = 0.05\,\mathrm{s}$, satisfying~\eqref{eq:dt-hocbf-bound}), $\lambda_1 = \lambda_2 = 1.0\,\mathrm{s^{-1}}$, $u_{\max} = 1.0\,\mathrm{m/s^2}$, a $6$-vertex polygonal robot footprint, and $1$--$6$ purely translating convex polygonal obstacles ($4$--$12$ vertices) moving at constant velocity unless stated otherwise. All scenes are drawn at random from a fixed seed ($\mathrm{seed}=0$) and run on an Intel Core i7-1260P CPU.

\subsection{Closed-Loop Navigation Performance}
\label{subsec:closed_loop}

We run $250$ closed-loop episodes ($50$ per family) across five scenarios: \emph{doorway} (a narrow gap between static posts), \emph{offset} (static obstacles obstructing the nominal path), \emph{crossing} (moving obstacles intercepting the path laterally), \emph{closing} (head-on approaching obstacles), and \emph{cut-in} (slower obstacles merging ahead along the path). A nominal PD controller, saturated to $\mathcal{U}$, tracks a straight start-to-goal line without a backup recovery controller; episodes terminate at the goal, at a $60\,\mathrm{s}$ timeout, or upon infeasibility.

We compare ELRH against two baselines following~\cite{trende2025lrh} and sharing our HOCBF formulation: Minimum Distance Hyperplane (\textbf{MDH}), which uses the closest-point normal, maximizing geometric separation, and \textbf{C-LRH} (continuous LRH), our implementation of the LRH approach, which solves~\eqref{eq:joint-problem} directly via IPOPT. Table~\ref{tab:closed_loop_250} reports the 50th, 80th, and 90th percentiles of completion time and cumulative intervention cost $J_{\mathrm{int}} = \sum_{k=0}^{K-1}\delta t\, J(\Delta u_k) = \sum_{k=0}^{K-1}\frac{\delta t}{2}\|\Delta u_k\|_2^2$, ranking unsuccessful runs last.

MDH times out in $17.2\%$ of episodes because normals maximizing separation block forward progress, notably in closing geometries. Across the 248 episodes completed by both methods, ELRH achieves lower $J_{\mathrm{int}}$ than C-LRH in $63.7\%$ of trials ($20.6\%$ ties, $15.7\%$ higher), improving both median and tail costs. While per-step optimality does not guarantee minimum cumulative intervention cost (early suboptimal deviations in C-LRH can occasionally steer into less constrained future states), ELRH strictly minimizes instantaneous intervention cost $J(\Delta u)$ at every step. We excluded the braking-distance variant of~\cite{trende2025lrh} because it caused frequent IPOPT convergence failures when $n_i^\top v_{\mathrm{rel},i}=0$ (non-differentiable).

\begin{table}[!t]
\centering
\caption{Closed-loop performance across $250$ multi-obstacle episodes: percentiles of completion time and cumulative intervention cost $J_{\mathrm{int}}$. Best values are in \textbf{bold}; dashes (`--') denote percentiles beyond the success rate (timeout or infeasibility).}
\label{tab:closed_loop_250}
\scriptsize
\renewcommand{\arraystretch}{0.92}
\setlength{\tabcolsep}{2.8pt}
\begin{tabular}{@{}ll rrr rrr@{}}
\toprule
& & \multicolumn{3}{c}{Time (s) $\downarrow$} & \multicolumn{3}{c}{$J_{\mathrm{int}}$ $\downarrow$} \\
\cmidrule(lr){3-5} \cmidrule(lr){6-8}
Family & Method & 50th & 80th & 90th & 50th & 80th & 90th \\
\midrule
Doorway & MDH & 8.60 & 8.65 & 8.65 & 1.023 & 1.059 & 1.074 \\
& C-LRH & \textbf{7.45} & \textbf{7.45} & \textbf{7.45} & \textbf{0.000} & \textbf{0.000} & \textbf{0.000} \\
& ELRH & \textbf{7.45} & \textbf{7.45} & \textbf{7.45} & \textbf{0.000} & \textbf{0.000} & 0.001 \\
\addlinespace[2pt]
Offset & MDH & 12.50 & \multicolumn{1}{c}{--} & \multicolumn{1}{c}{--} & 3.592 & \multicolumn{1}{c}{--} & \multicolumn{1}{c}{--} \\
& C-LRH & 8.10 & 8.31 & 8.45 & 0.940 & 1.091 & 1.161 \\
& ELRH & \textbf{8.03} & \textbf{8.16} & \textbf{8.30} & \textbf{0.927} & \textbf{1.024} & \textbf{1.054} \\
\addlinespace[2pt]
Crossing & MDH & 11.50 & 12.18 & 12.90 & 2.695 & 2.942 & 3.060 \\
& C-LRH & 7.40 & 9.35 & 9.46 & 0.640 & 2.164 & 2.338 \\
& ELRH & \textbf{7.35} & \textbf{8.10} & \textbf{9.21} & \textbf{0.586} & \textbf{1.568} & \textbf{2.201} \\
\addlinespace[2pt]
Closing & MDH & 13.80 & \multicolumn{1}{c}{--} & \multicolumn{1}{c}{--} & 3.733 & \multicolumn{1}{c}{--} & \multicolumn{1}{c}{--} \\
& C-LRH & 7.95 & 8.15 & 8.35 & 0.888 & 1.071 & 1.170 \\
& ELRH & \textbf{7.90} & \textbf{8.00} & \textbf{8.05} & \textbf{0.851} & \textbf{0.942} & \textbf{0.973} \\
\addlinespace[2pt]
Cut-in & MDH & 15.15 & 22.93 & 24.69 & 6.023 & 10.270 & 11.521 \\
& C-LRH & 8.32 & 8.90 & 9.21 & 1.128 & 1.606 & 1.776 \\
& ELRH & \textbf{8.20} & \textbf{8.36} & \textbf{8.52} & \textbf{1.022} & \textbf{1.202} & \textbf{1.410} \\
\midrule
\textbf{Overall} & MDH & 11.70 & 24.13 & \multicolumn{1}{c}{--} & 3.016 & 11.149 & \multicolumn{1}{c}{--} \\
& C-LRH & 7.95 & 8.36 & 8.85 & 0.852 & 1.162 & 1.639 \\
& \textbf{ELRH} & \textbf{7.90} & \textbf{8.15} & \textbf{8.35} & \textbf{0.821} & \textbf{1.024} & \textbf{1.192} \\
\bottomrule
\end{tabular}
\end{table}

\subsection{Global Minimum Recovery}
\label{subsec:one_step}

To illustrate Theorem~\ref{thm:global-optimality}, we evaluate a static two-obstacle doorway scenario (Fig.~\ref{fig:figure}(a)) at the state shown, measuring the per-step intervention cost $J_{\mathrm{step}} = J(\Delta u) = \frac{1}{2}\|\Delta u\|_2^2$. ELRH deterministically recovers the global optimum ($J_{\mathrm{step}} = 0.346$). In contrast, initializing C-LRH with either the MDH closest-point normal or the previous timestep's active normal traps IPOPT in a suboptimal local minimum ($J_{\mathrm{step}} = 0.415$, a $19.9\%$ higher cost) with a more restrictive safe set. Across $20$ random initial normal orientations (uniform in angle), C-LRH finds the global minimum in only $30\%$ of solves and converges to the local minimum in the remaining $70\%$.

\subsection{Computational Scalability}
\label{subsec:scaling}

We benchmark per-step runtime in an expanding-ring layout around the robot by scaling obstacle counts $N_{\mathrm{obs}} \in [1, 16]$ and uniform obstacle vertex counts $N_{z,i} = N_z \in [4, 12]$ ($\forall i$), comparing ELRH against MDH-initialized C-LRH and a discretized LRH baseline (\textbf{D-LRH}) extending~\cite{trende2025lrh}, which samples hyperplane angles and solves a convex program for each orientation combination. Since~\cite{trende2025lrh} restricts fine-grid discretization to single-obstacle scenes, we use a coarse grid ($M=5$ angles per obstacle) to contrast exponential $O(M^{N_{\mathrm{obs}}})$ enumeration with ELRH's polynomial search. Each $(N_{\mathrm{obs}}, N_z)$ configuration uses $20$ independent scenes, each timed as the median of $3$ runs.

As shown in Fig.~\ref{fig:figure}(b) ($N_z = 8$), D-LRH exceeds a $5\,\mathrm{s}$ timeout by $N_{\mathrm{obs}} = 6$, whereas even in a dense test of $N_{\mathrm{obs}} = 16$ simultaneous obstacles, ELRH achieves a median solve time of $41.8\,\mathrm{ms}$ over the $20$ scenes, a $6.1\times$ speedup over C-LRH ($253.0\,\mathrm{ms}$). Even exhaustive candidate evaluation without early termination (\textbf{ELRH exh.}) takes $55.3\,\mathrm{ms}$ at $N_{\mathrm{obs}} = 16$. Varying polygon complexity across $N_z \in [4, 12]$ (Fig.~\ref{fig:figure}(c)) alters runtime by less than $5\,\mathrm{ms}$ at $N_{\mathrm{obs}} \in \{4, 12\}$, confirming that $N_{\mathrm{obs}}$ dominates solve time and that detailed polygonal footprints add negligible overhead.

\section{Conclusion}
\label{sec:conclusion}

This work introduced ELRH, a solver-free framework for control-aware separating hyperplane selection in planar polygonal navigation. Exploiting Minkowski-difference geometry, ELRH reduces the non-convex joint optimization problem to algebraic evaluation over a polynomial candidate set of control interventions, achieving per-step global optimality under an explicit sampled-data safety guarantee. Compared to a nonlinear solver, ELRH lowered 90th-percentile cumulative intervention cost by $27.3\%$ across 250 episodes and achieved a $6.1\times$ runtime speedup at $16$ obstacles. Future work will address rotating bodies, multi-step horizon planning, and extension to 3D aerial navigation.


\bibliographystyle{IEEEtran}
\bibliography{references}

\end{document}